\documentstyle[psfig]{mn} 
\newif\ifAMStwofonts 
 
\def\kms{\thinspace\hbox{$\hbox{km}\thinspace\hbox{s}^{-1}$}}

\def\cc{\thinspace\hbox{$\hbox{cm}^{-3}$}}
\def\cm2{\thinspace\hbox{$\hbox{cm}^{-2}$}}
\def\ldo{$\lambda$}
\def\mpm{$\pm$}
  
\def\hb{\hbox{$\hbox{H}\beta$}}
\def\hg{\hbox{$\hbox{H}\gamma$}}
\def\hd{\hbox{$\hbox{H}\delta$}}

\def\msunyr{\thinspace\hbox{$\hbox{M}_{\odot}\thinspace\hbox{yr}^{-1}$}}
  
\def\kmsmpc{\thinspace\hbox{$\hbox{km}\thinspace\hbox{s}^{-1} \thinspace\hbox{Mpc}^{-1}$}}

\def\gapprox{$_>\atop{^\sim}$}     
\def\lapprox{$_<\atop{^\sim}$}        
\def\si{$\sim$}

\def\ldo{$\lambda$}
\def\arcdeg{\hbox{$^\circ$}}
\def\arcmin{\hbox{$^\prime$}}

\def\farcs{\hbox{$.\!\!^{\prime\prime}$}}

\def\hours{\hbox{$^{\rm h}$}}

\def\minutes{\hbox{$^{\rm m}$}}

\def\fseconds{\hbox{$.\!\!^{\rm s}$}}
\newcommand{\E}[1]{$\,10^{#1}$}                         
\newcommand{\ET}[1]{$\times 10^{#1}$}        
\newcommand{\zp}[1]{\left( {#1} \right)}               
\newcommand{\smc}[1]{\,\small \sc {#1}}

\ifoldfss 
  \ifCUPmtlplainloaded \else 
    \NewTextAlphabet{textbfit} {cmbxti10} {} 
    \NewTextAlphabet{textbfss} {cmssbx10} {} 
    \NewMathAlphabet{mathbfit} {cmbxti10} {} 
    \NewMathAlphabet{mathbfss} {cmssbx10} {} 
  \fi 
  \ifAMStwofonts 
    \ifCUPmtlplainloaded \else 
      \NewSymbolFont{upmath} {eurm10} 
        \NewSymbolFont{AMSa} {msam10} 
      \NewMathSymbol{\upi}     {0}{upmath}{19} 
      \NewMathSymbol{\umu}     {0}{upmath}{16} 
      \NewMathSymbol{\upartial}{0}{upmath}{40} 
      \NewMathSymbol{\leqslant}{3}{AMSa}{36} 
      \NewMathSymbol{\geqslant}{3}{AMSa}{3E}

    \fi 
  \fi 
\fi 
 
\ifnfssone 
  \newmathalphabet{\mathit} 
  \addtoversion{normal}{\mathit}{cmr}{m}{it} 
  \addtoversion{bold}{\mathit}{cmr}{bx}{it} 
  \newmathalphabet{\mathbfit} 
  \addtoversion{normal}{\mathbfit}{cmr}{bx}{it} 
  \addtoversion{bold}{\mathbfit}{cmr}{bx}{it} 
  \newmathalphabet{\mathbfss} 
  \addtoversion{normal}{\mathbfss}{cmss}{bx}{n} 
  \addtoversion{bold}{\mathbfss}{cmss}{bx}{n} 
  \ifAMStwofonts 
    \ifCUPmtlplainloaded \else 
      \UseAMStwoboldmath 
      \makeatletter 
      \new@mathgroup\upmath@group 
      \define@mathgroup\mv@normal\upmath@group{eur}{m}{n} 
      \define@mathgroup\mv@bold\upmath@group{eur}{b}{n} 
      \edef\UPM{\hexnumber\upmath@group} 
      \new@mathgroup\amsa@group 
      \define@mathgroup\mv@normal\amsa@group{msa}{m}{n} 
      \define@mathgroup\mv@bold\amsa@group{msa}{m}{n} 
      \edef\AMSa{\hexnumber\amsa@group} 
      \makeatother 
      \mathchardef\upi="0\UPM19 
      \mathchardef\umu="0\UPM16 
      \mathchardef\upartial="0\UPM40 
      \mathchardef\leqslant="3\AMSa36 
      \mathchardef\geqslant="3\AMSa3E 
    \fi 
  \fi 
\fi 
 
\ifnfsstwo 
  \DeclareMathAlphabet{\mathbfit}{OT1}{cmr}{bx}{it} 
  \SetMathAlphabet\mathbfit{bold}{OT1}{cmr}{bx}{it} 
  \DeclareMathAlphabet{\mathbfss}{OT1}{cmss}{bx}{n} 
  \SetMathAlphabet\mathbfss{bold}{OT1}{cmss}{bx}{n} 
  \ifAMStwofonts 
    \ifCUPmtlplainloaded \else 
      \DeclareSymbolFont{UPM}{U}{eur}{m}{n} 
      \SetSymbolFont{UPM}{bold}{U}{eur}{b}{n} 
      \DeclareSymbolFont{AMSa}{U}{msa}{m}{n} 
      \DeclareMathSymbol{\upi}{0}{UPM}{"19} 
      \DeclareMathSymbol{\umu}{0}{UPM}{"16} 
      \DeclareMathSymbol{\upartial}{0}{UPM}{"40} 
      \DeclareMathSymbol{\leqslant}{3}{AMSa}{"36} 
      \DeclareMathSymbol{\geqslant}{3}{AMSa}{"3E} 
    \fi 
  \fi 
\fi 
 
\ifCUPmtlplainloaded \else 
  \ifAMStwofonts \else 
    \def\upi{\pi} 
     
    \def\upartial{\partial} 
  \fi 
\fi 
 
 \title[GRB~010222]
{The host galaxy of GRB~010222: The strongest damped Lyman-$\alpha$ system known.\thanks{Based on 
observations made with the 4.2-m William Herschel Telescope, 
operated on the island of La Palma by the Isaac Newton
Group in the Spanish Observatorio del Roque de los Muchachos of the
Instituto de Astrof\'{\i}sica de Canarias.}}
\author[I. Salamanca, et al.]
       {I. Salamanca$^1$\thanks{e-mail: isabel@science.uva.nl}, L. Kaper$^1$, P. M. Vreeswijk$^1$, S. L. Ellison$^2$, E. Rol$^1$
     \newauthor N. Tanvir$^3$,  R.A.M.J. Wijers$^4$, P.F.L. Maxted$^5$ and E. van den Heuvel$^1$, \\
$^1$Astronomical Institute ``Anton Pannekoek'', University of Amsterdam, Kruislaan 403, 1098 SJ Amsterdam, The Netherlands \\
$^2$European Southern Observatory, Alonso de Cordova 3107, Vitacura, Casilla 19001, Santiago 19, Chile\\
$^3$Department of Physical Sciences, University of Hertfordshire, College Lane, Hatfield AL10 9AB, United Kingdom \\
$^4$Department of Physics \& Astronomy, SUNY, Stony Brook, New York 11794-3800, U.S. \\
$^5$Astrophysics Group, School of Chemistry \& Physics, Keele University, Staffordshire ST5 5BG,United Kingdom }

\date{Accepted ... 
      Received ...; 
      in original form: 2001 December 3} 
 
\pagerange{\pageref{firstpage}--\pageref{lastpage}} 
\pubyear{2001} 
 
\begin{document} 
 
\maketitle 
 
\label{firstpage} 
 
\begin{abstract} 
Analysis of the absorption lines in the afterglow spectrum of the
gamma-ray burst GRB~010222 indicates that its host galaxy (at a
redshift of z=1.476) is the strongest damped Lyman-$\alpha$ (DLA)
system known, having a very low metallicity and modest dust content. This
conclusion is based on the detection of the red wing of
Lyman-$\alpha$ plus a comparison of the equivalent widths of
ultraviolet Mg{\smc I}, Mg{\smc II}, and Fe{\smc II} lines
with those in other DLAs.  The column density
of H{\smc I}, deduced from a fit to the wing of Lyman-$\alpha$, is (5
$\pm$ 2)\ET{22} \cm2.  The ratio of the column
densities of Zn and Cr lines suggests that the dust content in our line of sight through the
galaxy is low. This could be due to either dust destruction by the ultraviolet
emission of the afterglow or to an initial dust composition different
to that of the diffuse interstellar material, or a combination of both.
\end{abstract}

\begin{keywords} 
Gamma-ray: burst - Galaxies: absorption lines
\end{keywords} 

\section{Introduction} 

Gamma Ray Bursts (GRBs), the most powerful cosmic explosions, are very important as probes of the
high-redshift universe. This is due to several factors: (i) GRBs occur
at cosmological distances, the most distant so far
being at z=4.5 (GRB~000131, Andersen et al. 2000), (ii) they can be
extremely bright and therefore could be seen (though for a very
short period of time) up to redshift 20 (e.g. Wijers et al.
1998; Lamb \& Reichart 2000), (iii) the burst and its afterglow illuminate the material
situated between the GRB and the observer, providing an opportunity to study galaxies that otherwise would be too faint to be observed 
spectroscopically. This is particularly relevant for the host galaxies of GRBs where, in analogy to quasar (QSO) absorption
line studies, the strengths of the absorption lines can be used to
measure important properties, such as their gas content, dynamical structure, and metallicity.
In addition, since they are transient phenomena, they do not cause
the large scale ``proximity effect'' of quasars, i.e. they do not modify their
environment at large distances, although they are expected to do so in their immediate
vicinity (see section~\ref{sec:discussion}), 
(iv) the currently most popular physical explanation of this phenomenon
is the so-called fireball model (cf. Rees \& Meszaros 1992) which is a
relativistic explosion caused by either the collapse of a massive star
(Woosley 1993; Paczynski 1998) or the merging of two compact objects
(Lattimer \& Schramm 1974). In the first case the rate of GRBs in the universe
should be directly linked to the instantaneous star formation rate (SFR). 
This indeed seems to be supported by the observations, since the host galaxies
of GRBs\footnote{Actually, this is true for bursts whose duration is longer that 2 seconds, since they are
the only ones which have been localized up to now.}
are undergoing strong star-formation activity
(e.g. Fruchter et al. 2001; Vreeswijk et al. 2001).
Therefore, GRBs offer the potential to measure the cosmic
star-formation rate (SFR) as a function of redshift.


In this paper we report on spectroscopic observations of the GRB~010222
afterglow taken \si 23 hours after the burst. We present evidence
that its host galaxy is a damped Lyman-$\alpha$
absorption (DLA) system with the strongest column density of neutral
hydrogen and the lowest metallicity currently known. This has strong
implications for both theories of the nature of GRBs and their host galaxies, and for the studies of DLA absorption
systems. In particular, it shows that DLA systems with N(H{\smc I}) $>
10^{22}$ \cm2 {\it do} exist, and that the way to detect them may be by looking at
GRB afterglow spectra rather than QSO
absorption line spectra. This is of extreme importance because DLAs are key components of the universe 
at high redshift due to the fact that, although relatively rare, they account for most of the neutral gas available for star formation. 

In section~\ref{sec:obs} we present the observations and describe the
calibration procedures.  In section~\ref{sec:analysis} the reduced
spectra are analysed, resulting in a detailed identification of the
absorption lines, and the determination of the metallicity and dust content of the
gas producing them. Finally, in section~\ref{sec:nature} we determine the
nature of the galaxy hosting GRB~010222 and compare it to the host
galaxies of other GRBs, as well as with other Mg{\smc II} and damped
Ly$\alpha$ absorption systems. The discussion and conclusions are presented in
section~\ref{sec:discussion} and section~\ref{sec:conclusions}, respectively.

\subsection{GRB~010222}

GRB~010222 was discovered on February 22, 2001, at 7:23:30 U.T. by the
Wide Field Camera (WFC) on board {\it Beppo}SAX (Piro 2001).  This
burst is one of the brightest bursts ever recorded, with a fluence of
(9.3 $\pm$ 0.3)\ET{-5} in the energy range from 40 to 700~keV.  Soon
after its detection, an optical transient was discovered with
coordinates RA = 14\hours 52\minutes 12\fseconds55 and DEC $=$+43\arcdeg 01\arcmin 06\farcs2 (J2000), 
and magnitudes V\si 18 and R $=$ 18.4\mpm 0.1 (Henden 2001; McDowell et al. 2001; Henden \& Vrba
2001). A series of optical spectra of the afterglow taken shortly after the alert allowed the
determination of the redshift of the host galaxy (Garnavich et
al. 2001; Jha et al. 2001; Bloom et al. 2001; Castro et al., 2001),
which is $z=1.476$. For a cosmology with {H$_0=65$ \kmsmpc,
$\Omega_m=0.3$, $\Lambda=0.7$}, the luminosity distance is D$_L =
$3.6\ET{28} cm, and the isotropic energy output is 7.8\ET{53} erg (in
't Zand et al. 2001), which makes GRB~010222 the third most energetic of all
GRBs with known redshift.

With regard to the host galaxy,
observations of GRB~010222 at mm- and sub-mm-wavelengths have led to
the interpretation that it is vigorously forming stars at
a rate of $\sim$500~M$_{\odot}$~yr$^{-1}$ (Frail et al. 2001), whilst in the optical and near-infrared
it is a faint (V$=$26.0\mpm 0.1) compact (FWHM\si 0\farcs15) galaxy (Fruchter et al. 2001).

\section{WHT spectra of the optical afterglow}
\label{sec:obs}
We obtained optical spectra of the afterglow of GRB~010222 on 2001 February
23, at 06:03:08.976 U.T. (i.e. 22.66 hours after the burst) with
the double spectrograph ISIS at the 4.2-m {\it William Herschel
Telescope} (WHT), of the Isaac Newton Group (ING) situated at ``El
Roque de los Muchachos'' observatory, La Palma, Spain. We used an EEV12 and a
TeK CCD (charged coupled device) as detectors, together with the B300B and R158R
gratings in the blue and red arm, respectively. The total exposure
time was 50 minutes.  The spectral resolution, as measured from the
width of the arc lines, is 3.3~\AA\ in the blue arm, and 5.8~\AA\ in
the red arm.  The nominal wavelength coverage for the blue arm is from
2728 to 6267~\AA. However, the camera optics in the blue arm suffer
from vignetting, resulting in a reduced effective wavelength range
(3042 to 5954~\AA).
The red arm covers the range from 6140 to 9115~\AA. The log of the
observations is listed in Table~\ref{tab:log}.  

\begin{figure*}
\vspace{0cm}  
\psfig{figure=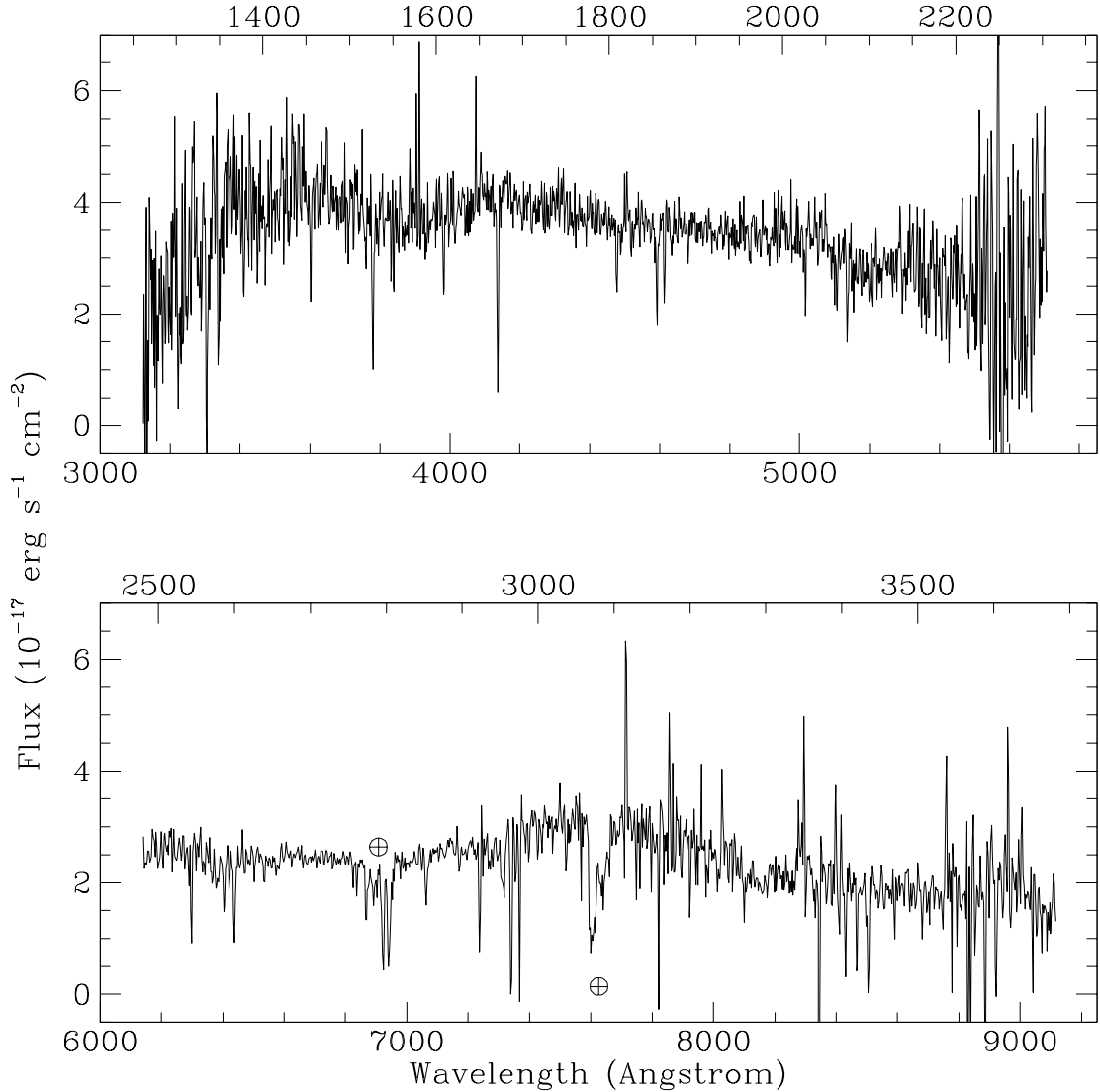,width=16cm} 
\caption{The optical spectrum of the afterglow of GRB~010222. For each panel, the observed wavelength is given at the bottom,
and at the top, the wavelength at the host galaxy rest-frame (i.e. corrected by red-shift z$=$1.476).
The spectrum includes many, mostly
saturated, absorption lines which can be identified with common
resonance lines at the redshift of the host galaxy ($z=1.476$) and at 
two other absorption systems at redshifts ($z = 1.156$ and 0.927).
The telluric absorption lines are marked with
$\oplus$.
At the redshift of the host galaxy, the center of
the Lyman-$\alpha$ line should be located at \si 3011~\AA, which is just
outside the wavelength range covered by our spectrum. However, we can see
its red wing extending up to 3400~\AA\ (observed wavelength). We do not detect
the interstellar graphite feature at 2175~\AA\ (rest-frame); the origin of the ``bump'' in the continuum around
7600~\AA\ is unclear.}
\label{fig:spectra}
\end{figure*}

\begin{figure*}
\vspace{0cm}  
\psfig{figure=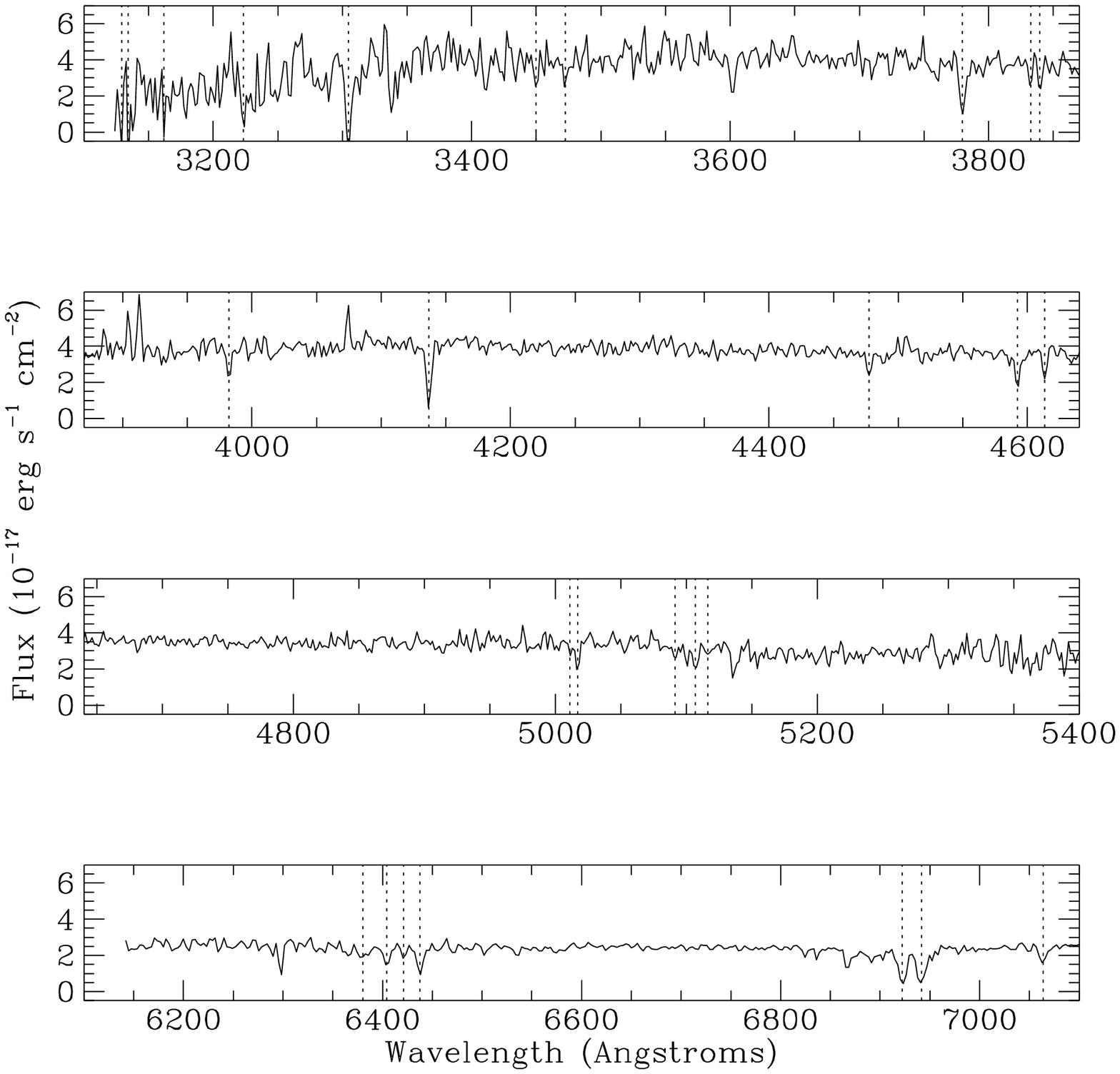,width=16cm} 
\caption{The optical spectrum of GRB~010222. We zoom in those areas which contain the lines listed in Table~\ref{tab:lineas_z1}
and we marked them with a dotted line. The signal-to-noise per resolution element is \si 13. }
\label{fig:spectra_zoom}
\end{figure*}

The data were debiased, divided by a normalized flat field and
extracted following standard procedures in {\smc IRAF}\footnote{{\smc
IRAF} is distributed by the National Optical Astronomy Observatories,
which are operated by the Association of Universities for Research in
Astronomy, Inc., under cooperative agreement with the National Science
Foundation.}. The spectra were then calibrated in wavelength using
Ar-Cu arc spectra. A comparison between the observed and theoric wavelength of the Balmer series in the standard star gives
an average error of 1.15 \AA\ in the blue spectrum, and 0.91 \AA\ in the red spectrum. Sky emission lines were removed by subtraction of
the average sky background spectrum next to the object. However, in the region redwards of 
\si 7600 \AA\ the sky substraction is difficult and remnant sky emission
lines are left.  This does not affect the subsequent analysis as we have not used this part of the 
spectrum in our analysis. Finally, the flux calibration was performed using
standard star BD+26$^{\circ}$2606. The atmospheric extinction was applied using the
mean extinction curve for La Palma. We present the final spectrum in Figs.~\ref{fig:spectra} and \ref{fig:spectra_zoom}.

\begin{table*}
\begin{minipage}[ctb]{95mm}
\caption{Log of the spectroscopic observations of the afterglow of GRB~010222. The data were taken on 2001 February 23, about
22.66 hours after the burst. The total exposure time is 50 minutes.}
\label{tab:log}
\begin{tabular}{cccc}
\hline
Instrument  & CCD/Grism & Spect. range  & Spect. resol.$^a$ \\
            &           & (\AA)    & (\AA\ / \kms)  \\
\hline
\hline
 ISIS Blue arm & EEV12 / B300B & 3042 - 5954 & 3.3 / 220 \kms  \\
 ISIS Red arm  & TeK / R158R   & 6140 - 9115 & 5.8 / 230 \kms \\
\hline 
\end{tabular}

$^a$As measured from the FWHM of the arc lines\\
\end{minipage}
\end{table*}

\section{Analysis of the absorption lines}
\label{sec:analysis}

\subsection{Line identification}
\label{sec:lines}

We identified several absorption lines in the spectrum of the
GRB~010222 afterglow, belonging to (at least) three different
absorption systems at redshifts z$=$1.476, z$=$1.156 and z$=$0.927
(Jha et al. 2001; Masetti et al. 2001; Castro et al. 2001).
Actually, the GRB~010222 afterglow is one of the richest in
terms of line detections; a compilation of absorption lines detected
in GRB afterglow spectra can be found in Vreeswijk et al. (2001).  In
this paper we will concentrate on the system at the highest redshift,
which we will adopt as belonging to the host galaxy of GRB~010222. 

The host galaxy's absorption lines and their identification are
compiled in Table~2, together with the rest-frame equivalent width (EW)
and some atomic parameters (the oscillator strength, {\it f}, and the
ionization potential of the ion, IP, in eV). For comparison, we have listed
the EW published by Jha et al. (2001) and Masetti et al. (2001). The
lines that we could not identify (neither at redshift 1.476, nor at the other two redshifts) are listed in Table~3.

It is important to determine whether the lines are saturated or not,
which is difficult given the low resolution of our spectrum.  Most of
the absorption lines are expected to have a doppler broadening less than
a few tens of \kms, which is much less than the spectral resolution ($\sim
220$~\kms).  Therefore, we have compared the observed ratio of the
line intensities to the theoretical one, which is given by the ratio
of their oscillator strengths. 
Furthermore, the absorption lines with an equivalent width larger than \si 0.5 \AA\ 
are most likely saturated (see for
example Pettini et al. 2001).  Accordingly, most of the absorption
lines detected in the afterglow spectrum of GRB~010222 are actually
saturated.

\begin{table*}
\begin{minipage}[ctb]{170mm}
\caption{Absorption lines identified at $z=1.476$. The first column
lists the observed wavelength, the second the identification, the
third the redshift, and the fourth the equivalent width in the
rest-frame of the host galaxy, i.e. divided by (1+z). The error on the
equivalent width is estimated by varying the location of the continuum. The error in the wavelength position is \si 1.15 \AA\ for the
blue spectrum and 0.91 \AA\ for the red spectrum. For
comparison, we list in the fifth and sixth columns the equivalent width published by
Jha et al. (2001) and by Masetti et al. (2001).  The next columns are
the oscillator strength of the transition, $f$ (taken from Bergeson \& Lawler 1993; Morton 1994, 2001; Savage \&
Sembach 1996; Bergeson et al. 1996; Vernor 1996 and Schectman et al. 1998), and the ionization potential of the corresponding ion,
IP, in eV. Finally, in the last column, we mention whether the line is as
well detected in absorption line systems towards quasars
(e.g. Jannuzi et al. 1998), and two supernovae, SN~1987A (Blades et
al. 1988) and SN~1993J (Marggraf \& Boer 2000).}
\label{tab:lineas_z1}
\begin{tabular}{lllllllll}
\hline
Obs. \ldo & Identification & Redshift & Eq. Width  & EW Jha   & EW Mas   & ~$f$ & IP & detected as well \\
~~(\AA)   &                &         & ~~~~~(\AA)& (+4.9 h) & (+1day)    &      & eV &        \\
\hline
\hline
3129.2 & Si{\smc II} 1260.42  & 1.483 & 1.54 \mpm 1.18 & 	     & 		   & 1.007000 & 8.151  & qso, 87A, 93J\\ 
3134.4 & Si{\smc II$^*$} 1264.74  & 1.478 & 1.74 \mpm 0.20 & 	     & 		   & 0.903400 & 8.151  & 87A \\	
3161.9 & C{\smc I} 1277.24    & 1.475 & 0.98 \mpm 0.07 & 	     & 		   & 0.096650 & 0      & 87A  \\ 	 
3223.3 & O{\smc I} 1302.17    & 1.475 & 0.94 \mpm 0.03 & 	     & 		   & 0.048870 & 0      & qso, 87A, 93J \\
3304.7 & C{\smc II*} 1334.53$^a$   & 1.476 & 2.34 \mpm 0.03 & 	     &  	   & 0.1278 & 11.26  & qso, 87A  \\ 	 
3449.8 & Si{\smc IV} 1393.76  & 1.475 & 0.67 \mpm 0.04 & 	     & 		   & 0.5280 & 33.49  & qso, sn87A, 93J \\
3472.3 & Si{\smc IV} 1402.77  & 1.475 & 0.44 \mpm 0.01 & 	     & 		   & 0.2620 & 33.49  & qso, 87A, 93J \\
3779.7 & Si{\smc II} 1526.71  & 1.476 & 1.65 \mpm 0.17 &1.4 \mpm 0.4 & 1.2 \mpm 0.3& 0.12700 & 8.151  & qso, 87A, 93J \\
3832.4 & C{\smc IV} 1548.20   & 1.475 & 0.52 \mpm 0.10 & 	     & 		   & 0.190800 & 47.89  & qso, 87A, 93J \\ 
3839.5 & C{\smc IV} 1550.78   & 1.476 & 0.71 \mpm 0.14 &1.9 \mpm 0.4$^f$ & 2.3 \mpm 0.2& 0.09522 & 47.89  & qso, 87A, 93J \\ 
3982.3 & Fe{\smc II} 1608.45$^b$  & 1.476 & 0.67 \mpm 0.15 & 	     & 		   & 0.0580 &  7.87  & qso, 87A \\
4136.9 & Al{\smc II} 1670.79  & 1.476 & 1.49 \mpm 0.09 &1.1 \mpm 0.3 & 1.1 \mpm 0.3& 1.800 & 5.986  & qso, 87A \\  
4477.4 & Si{\smc II} 1808.01$^c$  & 1.476 & 0.56 \mpm 0.02 &0.5 \mpm 0.3 & 0.7 \mpm 0.3& 0.00218& 8.151& qso, 87A, 93J \\
4592.2 & Al{\smc III} 1854.72$^d$ & 1.476 & 0.73 \mpm 0.09 &  	     & 		   & 0.5390 & 18.83  & qso, 87A, 93J \\ 
4613.3 & Al{\smc III} 1862.79 & 1.477 & 0.72 \mpm 0.07 & 	     & 		   & 0.2680 & 18.83  & qso, 87A, 93J \\ 
5011.2 & Zn{\smc II} 2026.14  & 1.473 & 0.21 \mpm 0.24 & 1.0 \mpm 0.3& 0.6 \mpm 0.2& 0.489000 & 9.394  & qso, 87A, 93J \\
5017.0 & Mg{\smc I} 2026.48   & 1.476 & 0.62 \mpm 0.02 & 	     & 		   & 0.1120	& 0 & qso, 87A, 93J \\   
5091.4 & Cr{\smc II} 2056.25  & 1.476 & 0.22 \mpm 0.07 & 	     & 		   & 0.1050    & 6.766  & qso, 87A \\
5107.1 & Zn{\smc II} 2062.66  & 1.476 & 0.31 \mpm 0.03 &0.7 \mpm 0.2 & 1.1 \mpm 0.2& 0.256000 & 9.394  & qso, 87A \\  
       & +Cr{\smc II} 2062.23 &       &                & 	     & 		   & 0.0780 & 6.766  & qso, 87A  \\   
5116.4 & Cr{\smc II} 2066.16$^e$  & 1.476 & 0.38 \mpm 0.33 & 	     & 		   & 0.05150 & 6.766  & 87A  \\
6380.2 & Mn{\smc II} 2576.88  & 1.477 & 1.21 \mpm 0.40& 0.7 \mpm 0.2 &0.5 \mpm 0.2 & 0.3508  & 7.435  & qso, 87A   \\  
6404.4 & Fe{\smc II} 2586.65  & 1.476 & 1.42 \mpm 0.68& 1.5 \mpm 0.2 &0.8 \mpm 0.3 & 0.069125 & 7.87   & qso, 87A, 93J \\  
6421.4 & Mn{\smc II} 2594.50  & 1.476 & 0.57 \mpm 0.38& 0.7 \mpm 0.2 &0.8 \mpm 0.3 & 0.2710  & 7.435  & qso, 87A   \\ 
6437.7 & Fe{\smc II} 2600.17  & 1.476 & 2.16 \mpm 0.70& 1.9 \mpm 0.2 &1.9 \mpm 0.2 & 0.2390  & 7.87   & qso, 87A, 93J \\
6922.3 & Mg{\smc II} 2796.35  & 1.476 & 3.33 \mpm 1.27   &3.0 \mpm 0.2 &2.8 \mpm 0.2 & 0.6123  & 7.646  & qso, 87A, 93J \\
6941.7 & Mg{\smc II} 2803.53  & 1.476 & 3.82 \mpm 1.65   &2.7 \mpm 0.2 &3.4 \mpm 0.3 & 0.3054  & 7.646  & qso, 87A, 93J \\
7063.6 & Mg{\smc I} 2852.96   & 1.476 & 1.32 \mpm 0.89   &0.9 \mpm 0.2 &1.0 \mpm 0.3 & 1.81000   &  0	& qso, 87A \\
\hline
\hline
\end{tabular}

$^a$ The line of C{\smc II} 1334.53 \AA\ is broad (155 \kms\ after
correction for the instrumental profile). It is also observed as a
broad feature in the spectrum of SN~1987A (Blades, Wheatley, Panagia et al. 1998).\\
$^b$ The line of Fe{\smc II} 1608.45 \AA\ is identified by Masetti et al.
(2001) as Zn{\smc II} 2062 \AA\ at $z=0.926$, but in this case we
should observe Zn{\smc II} 2026 \AA\ as well, which we do not.\\
$^c$ The Si{\smc II} 1808.01 \AA\ line should be accompanied by another one
at 1816.93 \AA, but the latter is contaminated by a sky emission line.\\
$^d$ Note that this line is identified by Masetti et al. (2001) as being Fe{\smc II} 2383 \AA\ at
$z=$0.926. \\
$^e$ This line is too strong which makes its identification uncertain. Maybe this line is blended with some other line. \\
$^f$ The separate doublet components are not resolved, hence the equivalent width corresponds to the sum of
both components.
\end{minipage}
\end{table*}

\begin{table*}
\begin{minipage}[ctb]{46mm}
\caption{Unidentified lines. The equivalent width is as measured in
the observer's frame.}
\label{tab:lineas_z?}
\begin{tabular}{ll}
\hline
Observed \ldo &  Equivalent Width  \\
~~~~~(\AA)   &   ~~~~~ (\AA)     \\
\hline
\hline 
3138.3 & 2.69 \mpm 0.18  \\
3178.5 & 2.32 \mpm 0.07  \\
3410.9 & 2.27 \mpm 0.31  \\
3709.8 & 0.78 \mpm 0.03 \\
4389.6 & 0.57 \mpm 0.17 \\
4488.6 & 0.53 \mpm 0.07  \\
4518.1 & 0.58 \mpm 0.03  \\
4525.2 & 0.32 \mpm 0.00  \\
4633.9 & 1.49 \mpm 0.26  \\
4681.2 & 0.63 \mpm 0.01  \\
4856.7 & 0.91 \mpm 0.46$^a$  \\
4859.6 & (sum)         	 \\
4914.0 & 0.45 \mpm 0.03 \\
5100.3 & 0.55 \mpm 0.02 \\
7340.1 & 7.02 \mpm 0.11 \\
7519.9 & 1.59 \mpm 0.27 \\
8504.5 & 7.63 \mpm 0.37 \\        
\hline
\end{tabular}

$^a$ This line is blended with the line at 4859.6 \AA.\\
\end{minipage}
\end{table*}

\subsection{Detection of Lyman-$\alpha$ at z$=$1.476}

\begin{figure*}
\vspace{0cm}  
\hbox{\psfig{figure=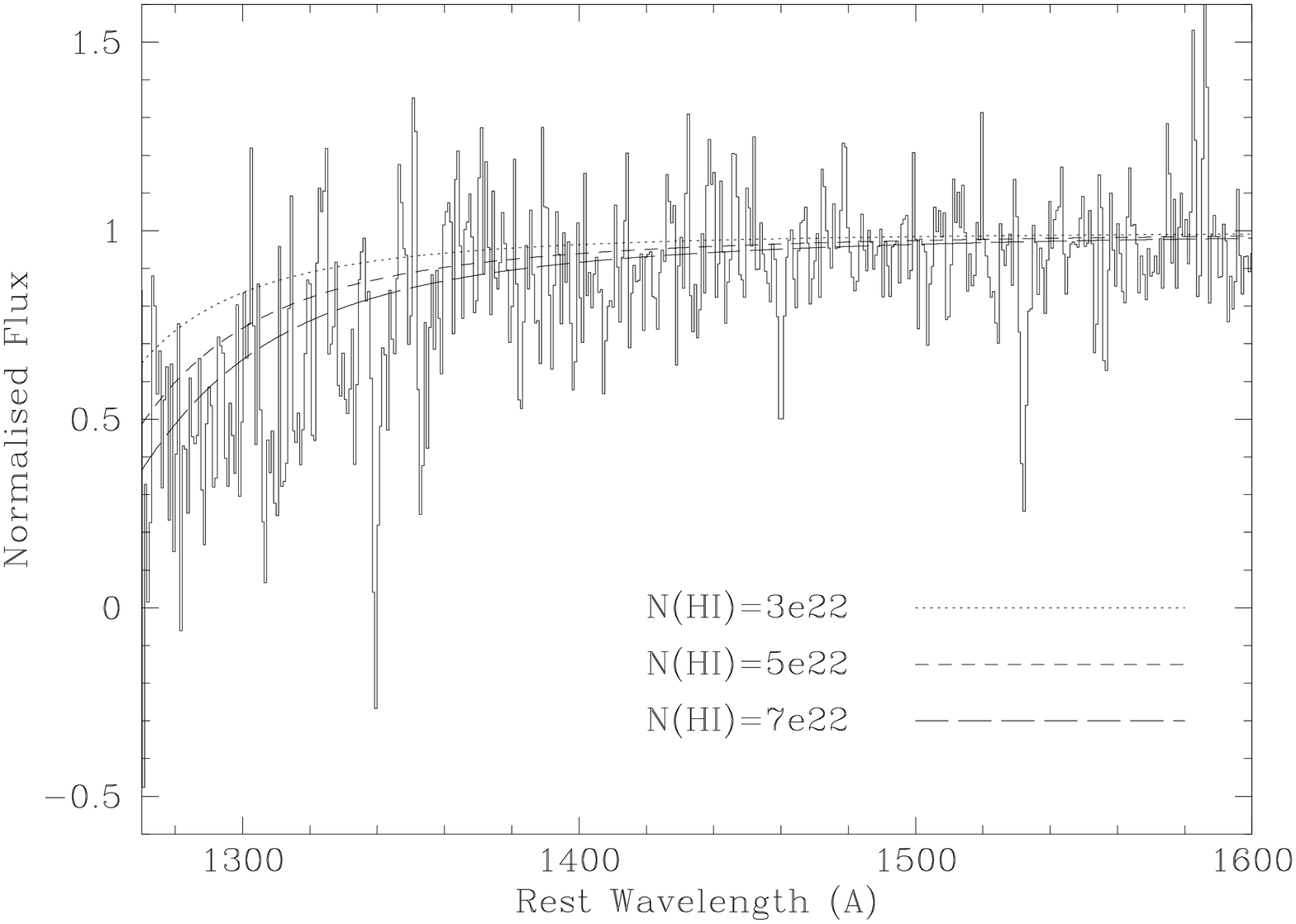,height=8cm,width=8cm,angle=-90}
\psfig{figure=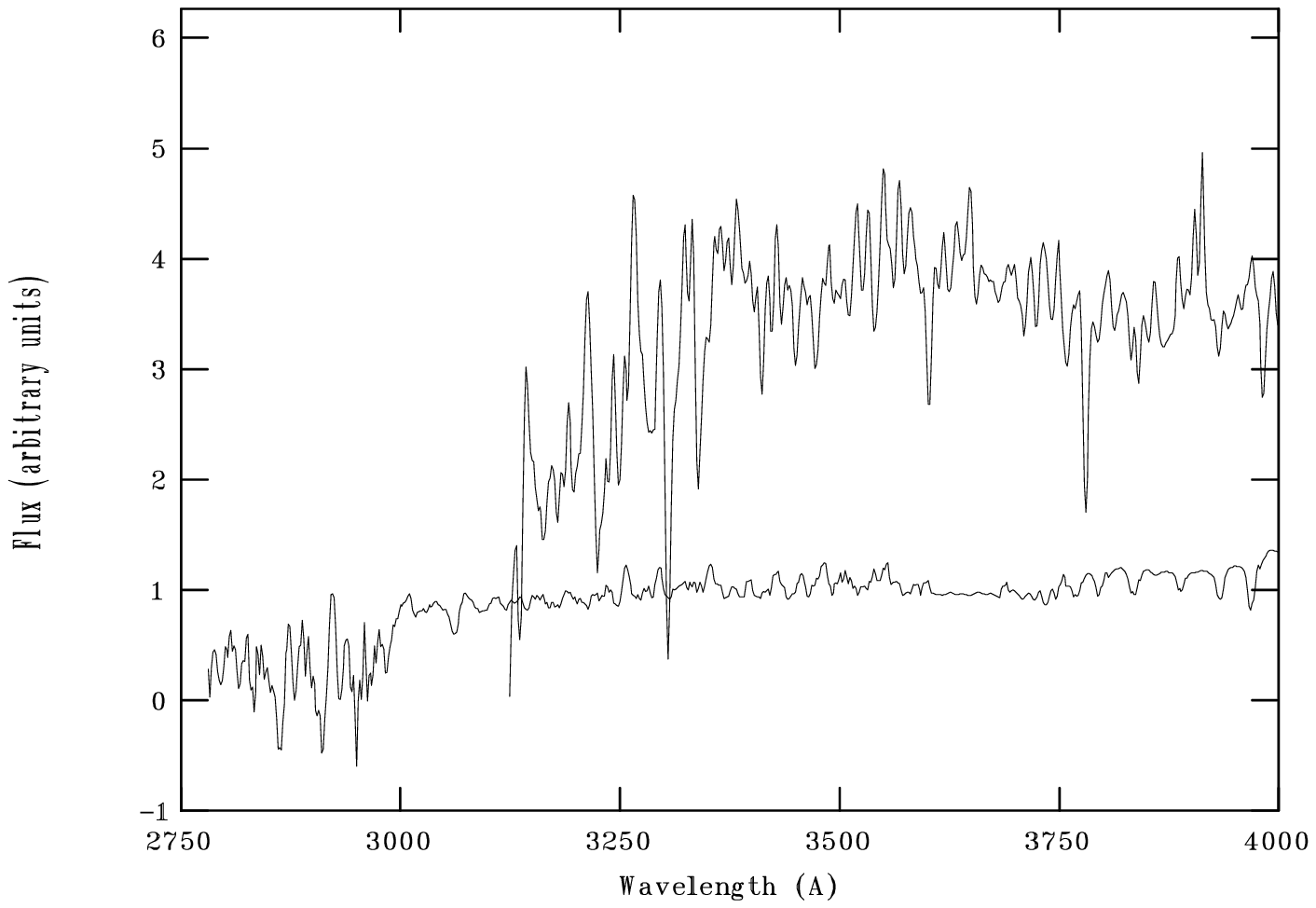,width=8.5cm,height=8cm}}
\vspace{-0.5cm}
\caption{Evidence that the host galaxy of GRB~010222 is a damped
Lyman-$\alpha$ system.  {\bf Left:} Our blue spectrum clearly shows
the red wing of Lyman-$\alpha$, which is matched with three Voigt
profiles corresponding to a column density of neutral hydrogen of 3\ET{22}, 5\ET{22} and 7\ET{22} \cm2. 
Before the fitting, the spectrum has been corrected for redshift and the adjacent continuum normalized to
unity.  {\bf
Right:} The spectrum of GRB~010222 (top) is shown next to that of the
standard star (bottom); the feature that we identify as the red wing
of Lyman-$\alpha$ is not present in the spectrum of the standard
star. Both spectra have been multiplied by an arbitrary constant so
that they can be displayed in the same panel, and they have been
binned so that their general trend is more obvious.}
\label{fig:lya}
\end{figure*}

The strongest line in the spectrum of GRB~010222's afterglow is actually Lyman-$\alpha$,
of which we can see its red wing  extending up to
\si 3400~\AA\ (observer's frame) or \si 1370 \AA\ (rest-frame).
The center of Lyman-$\alpha$ should be at 3010.8~\AA\ in the observer's frame, which is
just outside the wavelength region covered by our spectrum, close to the atmospheric
and instrumental cut-off.
We have checked that this feature is not
due to instrumental effects, such as flat fielding or vignetting. For
this purpose, we have compared the spectrum of GRB~010222 afterglow to that of the standard star, which was
observed right after the exposure of GRB~010222 was finished. We can
see in Fig.~\ref{fig:lya} that the feature that we identify as the red wing of Lyman-$\alpha$ is absent in the 
standard star spectrum and that the vignetting starts at
wavelengths shorter than \si 3000~\AA\ (observer's frame) or \si
1210~\AA\ (rest-frame).
This is what we expect according to the technical
documentation published by the ING group (Garcia Lorenzo 2000).
We have multiplied both spectra by an arbitrary constant, so they can
be plotted in the same diagram, and binned them so that their general
shape is more clear.

Consequently, we are confident that the feature observed at the blue
end of the spectrum of GBR~010222's afterglow {\it is} the red wing of
the damped Lyman-$\alpha$ absorption line of the host galaxy.

\subsection{Measuring the column densities}

The average redshift, based on the metal lines, of the absorption-line
system with the highest redshift is 1.467 (we have not taken into
account the blends of Zn{\smc II}$+$Cr{\smc II}).  According to Castro
et al. (2001) there is some substructure in the gas forming these
lines, with a separation of 106~\kms. The spectral resolution of our spectrum does not
enable us to resolve this, and therefore we can only deduce the global
properties of the galaxy or region where the GRB has taken place.

To measure the equivalent width, we used the routine {\tt splot}
within {\sc iraf}. The line was integrated with respect to the local
continuum, by just summing the flux over a given wavelength
interval. No gaussian fitting was done, except in the few cases where
the lines were partially blended, or the local continuum was noisy.
In these cases, we checked the obtained EW by fitting a single
Gaussian to the lines, yielding consistent results. To estimate the
error on the EW measurement, we varied
the position of the local continuum. The final value of the EW is the average of all measurements,
while the error is conservatively taken as the difference between the
highest and the lowest values obtained. The results are given in
Table~\ref{tab:lineas_z1}, where we have listed the EW published by Jha
et al. (2001) and Masetti et al. (2001) too.  In most cases, the three
independently obtained values are consistent within the errors. However, in
two cases the values do not agree: (a) the C{\smc IV} doublet at
1548, 1550~\AA\, which can be due to the fact that we resolve it,
while Jha et al. (2001) and Masetti et al.  (2001) do not; (b) the
Zn{\smc II} + Cr{\smc II} 2026 blend, but again this can be explained
by a difference in the spectral resolution: if we add the
equivalent width of Mg{\smc I} 2026 \AA\ line to that of the
Zn/Cr blend, then the values agree well.

\subsubsection{H{\smc I} column density}

We have deduced the column density of H{\smc I} by fitting a Voigt
profile to the red wing of Lyman-$\alpha$.  For that purpose, we
corrected the spectrum for redshift, and normalized the continuum to
unity.  Subsequently, we produced model profiles of damped absorption
lines, for three values of N(H{\smc I}): 3\ET{22}, 5\ET{22} and
7\ET{22} \cc, and compared them with the observed
Lyman-$\alpha$ wing. Even though the observed spectrum is noisy and
only part of the line profile is covered, a large value of N(H{\smc
I}) is required to match the observed red damping wing.  The value we
have adopted is N(H{\smc I}) = (5\mpm 2) \ET{22} \cc. This is formally (i.e. within the errors)
consistent with the value deduced from X-ray observations, (2.5\mpm 0.5)\ET{22} cm$^{-2}$ (in 't Zand et al. 2001).



\subsubsection{Column density of the other lines}

As we have discussed earlier,
most of the other lines seem to be saturated, and are, therefore, not suited to
calculate their column density. 
We limit ourselves to those lines that
are very likely {\it not} saturated. These are the Zn{\smc II} and
Cr{\smc II} lines, given their intrinsic low cosmic abundance and the
small value of the oscillator strength. Possibly the lines of Fe{\smc
II} 1608 \AA, and Mn{\smc II} 2594 \AA\ are unsaturated too, but nevertheless, we
regard the abundances determined from these transitions as lower
limits.

In order to deduce the equivalent widths of the individual lines in
the Zn + Cr blend, we start by measuring the EW of Zn{\smc II} 2026~\AA\
which is mainly dominated by Zn$^+$, and the EW of Cr{\smc II} 2056~\AA\ 
which is not blended. The total EW of Zn{\smc II} + Cr{\smc II}
2026~\AA\ is 0.21, but the contribution of Cr{\smc II} 2026 \AA\ can be
neglected, because the ratio of the oscillator strengths of Cr{\smc
II} 2056 \AA\ to Cr{\smc II} 2026 \AA\ is \si 22. Therefore, we assume that the
EW of Zn{\smc II} 2026 \AA\ is 0.21~\AA. This implies that the EW of
Zn{\smc II} 2062 \AA\ is \si 0.11 \AA. Since the total EW of the
Zn{\smc II} + Cr{\smc II} 2062 \AA\ blend is 0.31 \AA, then the EW of the
Cr{\smc II} 2062 \AA\ becomes \si 0.2 \AA.  The EW of the Cr{\smc II}
2066~\AA\ line has a very large error (87\% relative error).
Therefore, this line has not been taken into account to deduce the
column density of Cr$^+$.

Once we know the EW, we calculate N$_{\rm col}$ using the following
relationship, valid in the limit of weak lines, between the
(rest-frame) equivalent width and the column density (Scheffler \&
Elsasser 1987):
\[ N_{\rm col} =  1.131\times10^{12} \times \frac{\rm EW}{\lambda^2}
\frac{1}{f} \, \, \, \, \, \, {\rm cm} , \]
where $f$ is the oscillator strength of the transition, and $\lambda$ 
the wavelength in cm.

In the subsequent analysis, we have taken for each ion (Zn{\smc II}, Cr{\smc II},
Fe{\smc II}, and Mn{\smc II}) the average N$_{\rm col}$ of
the individual transitions, if more than one unsaturated line 
could be measured. We assume that the abundance of each element X is closely approximated
by X$^+$ because the second ionisation potential is above 13.6 eV.
 The results are summarized in Table~\ref{tab:abund}
and discussed in the following sections.


\section{The host galaxy of GRB~010222: \\
a damped Lyman-$\alpha$ system}
\label{sec:nature}


An absorption system is classified as a damped Lyman-$\alpha$ (DLA)
system if the colunm density of neutral hydrogen is larger than
2\ET{20} \cm2.  The column density of H{\smc I} in the host galaxy of
GRB~010222 is N(H{\smc I}) $=$ (5\mpm 2) \ET{22} \cm2.  This, {\it by
definition} makes the host galaxy of GRB~010222 a damped
Lyman-$\alpha$ system. Further support to the classification of the
host galaxy as a DLA comes from the analysis of Mg{\smc I}, Mg{\smc
II}, and Fe{\smc II} lines.  According to Rao \& Turnshek (2000), if
the absorption system has EW(Mg{\smc II} 2796) $>$ 0.6 \AA\ and
EW(Fe{\smc II} 2600) $>$ 0.5, it has 50\% probability of being a DLA
system. This is obviously the case for GRB~010222, as already pointed
out by Masetti et al. (2001).  Furthermore, all DLA systems in the Rao
\& Turnshek sample have Mg{\smc II} 2796, 2803 \AA\ doublet ratios
below 1.5, as is the case for GRB~010222
(Fig.~\ref{fig:ratioMgII}). The relationship between EW(Mg{\smc II}
2796) versus EW(Fe{\smc II} 2600) also indicates that the host galaxy
of GRB~010222 is a DLA: On the one hand both measured EWs are larger
than 0.6~\AA, and on the other hand their ratio is \si 1.5, very close
to the average value of 1.45 deduced by Rao \& Turnshek (2000)
(Fig.~\ref{fig:ratioMgII}).


Furthermore, the equivalent widths of other low ionization lines in
the spectrum of GRB~010222 are quite large compared to those of
absorption-line systems present in QSO spectra.  For example, the EW
of Si{\smc II} 1526~\AA\ is 1.65 \mpm 0.17~\AA, while the values
measured in the sample of Steidel \& Sargent (1992) range from 0.02
to 1.87~\AA.  Equally strong are the lines of Zn{\smc II} 2026,2062
\AA\ and Cr{\smc II} 2026,2056,2062 \AA. If we compare for example
their EW with those of the data collected by Pettini and collaborators
(1994, 1997 and 1999), we see, again, that GRB~010222 has the
highest values (even though it has the lowest metallicity, see later).

In order to estimate the metallicity and dust-to-gas ratio we compare
the abundances of elements that are known to be little depleted onto
dust to that of refractory elements that are easily depleted on to dust
grains. The best elements for such purpose are Zn (not depleted) and
Cr, Mn, Fe and Ni (large depletion).  Therefore, Zn is a good
indicator of the metallicity, whilst Cr, Mn, Fe and Ni can be used to
measure the dust content (for a more extended analysis of why
these elements are useful to study the metallicity and dust content, see
for example Pettini et al. 1990).

We adopt the usual notation, i.e.
\begin{equation}
 [X/H] = \log{\frac{N_{\rm col}(X)}{N_{\rm col}(H)}}_{\rm GRB} - \,\,\,\, 
\log{\frac{N_{\rm col}(X)}{N_{\rm col}(H)}}_\odot \, = \, \log{\delta(X)} \, , 
\end{equation}
where $\delta(X)$ is the depletion of element X, or in other
words, the fraction of element X in gas form, relative to its solar meteoritic abundance.
The latter values have been taken from Savage \& Sanders (1996). 

For the zinc abundance, and thus a measure of the enrichment of the
gas by metals, in the host of GRB~010222 we find a value of [Zn/H] =
-2.3 \mpm 0.5, indicating that the host galaxy has a very low
metallicity, about 200 times less than the solar fractional Zn
abundance. If we adopt N(H{\smc I}) = 2.5\ET{22} \cm2 (in 't Zand et al.
2001) then [Zn/H] becomes -2.01, about a factor 100 below solar,
thus the host galaxy would still be amongst the most metal poor DLAs
known. The lowest metallicity claimed up to now in DLAs is [Zn/H] =
-2.07 \mpm 0.1 (Molaro et al. 2001).

The dust content of the host galaxy can be deduced by comparing the Zn
abundance to the abundance of some refractory elements:
\begin{equation}
 [X/Zn] =log(\frac{\delta(X)}{\delta(Zn)}) 
\approx log\zp{\frac{f_{{\rm gas}}}{f_{{\rm gas}} + f_{{\rm dust}}}}
\end{equation}
where X is Cr, Fe or Mn, $\delta(X)$ its gas-phase abundance, and
$f_{{\rm gas}}$ and $f_{{\rm dust}}$ are the fraction of element X in gas and dust
respectively.  The value of [X/Zn] = -0.3 corresponds to the case
where the fraction of X in dust is equal to the fraction of X
contained in gas form ($f_{{\rm gas}} = f_{{\rm dust}}$).  The results are
summarized in Table~\ref{tab:abund}.  The abundances of Zn, Cr and Mn
are consistent with each other, while that of Fe is slightly smaller.

A few words of caution before reaching any conclusion: The iron line
Fe{\smc II} 1608~\AA\ has an EW slightly larger than 0.5~\AA, and
therefore could be saturated, so its abundance should be regarded as a
lower limit.  The same applies to Mn, which in addition has a
nucleosynthetic pattern that is not well understood, so we may be
seeing effects of nucleosynthesis, and not of depletion on dust
(Pettini et al. 2000).  Also, it is important to realize that there
might be effects due to the ionization of the gas caused by the local UV
radiation field of OB stars and/or of the GRB itself.  These effects
would raise the value of [Zn/Cr] which could mimic the absence of dust
(Howk \& Sembach 1999).  However, further observational proof of the
lack of dust in the host of GRB~010222 is the absence of the
``2175~\AA\ dip'', which is usually attributed to graphite dust grains
and detectable down to A$_V$ \si\ 0.1~mag (see also Jha et al. 2001 and
Masetti et al. 2001).

\begin{figure*}
\vspace{0cm}  
\hbox{\psfig{figure=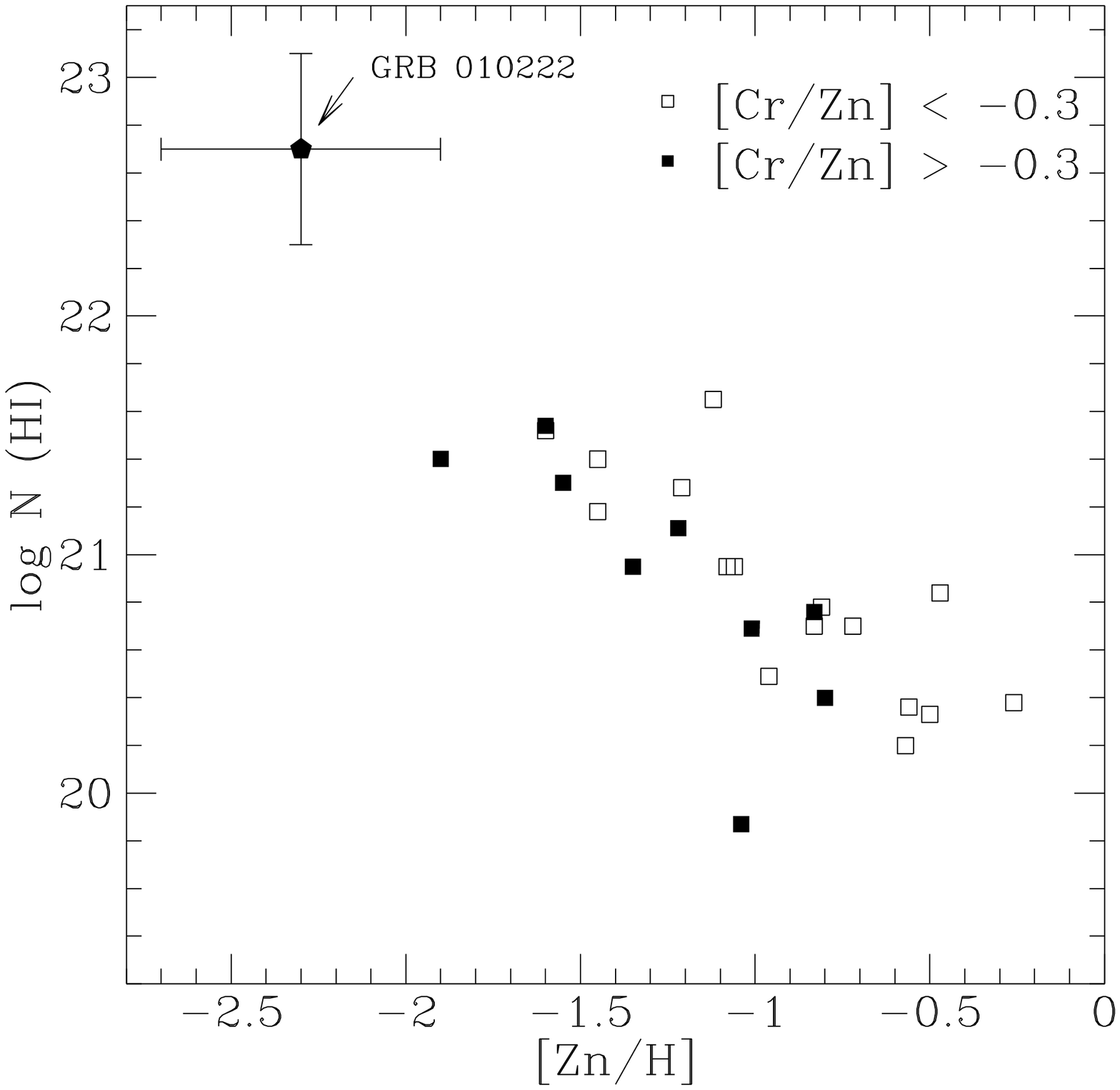,width=8cm}
\psfig{figure=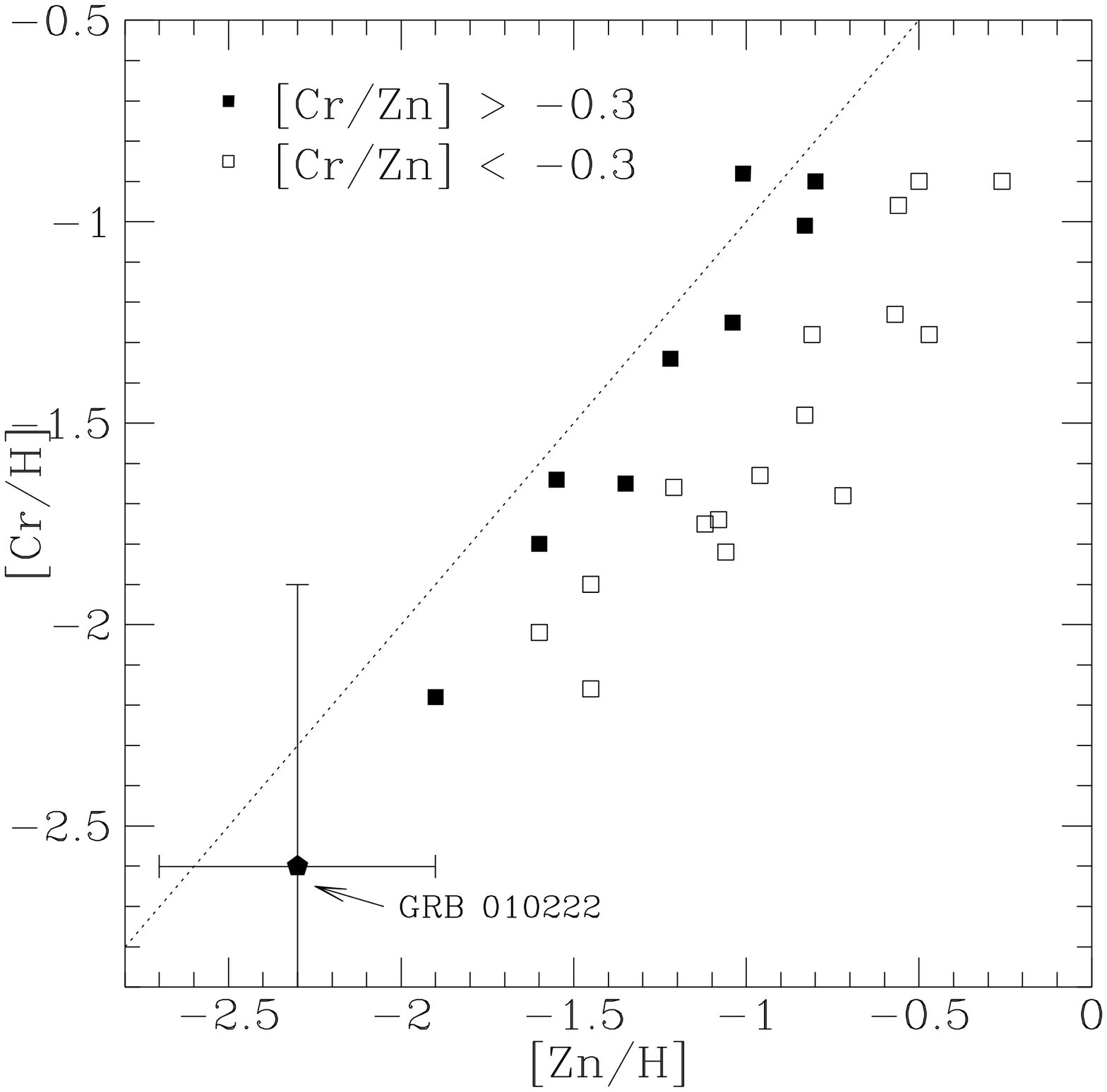,width=8cm}}
\caption{The host galaxy of GRB 010222 has the highest column density of 
H{\smc I} and the lowest metallicity of the whole sample of DLAs. 
The data of the latter have been taken from Pettini et al. (1994,
1997 and 1999). The values of the abundances of Zn and Cr and of
the column density of H{\smc I} are what one would expect from an
extrapolation of the correlations seen in both diagrams. A word of caution however: the
correlation between N(H{\smc I}) and [Zn/H] is thought to be due to selection
effects; there is no physical reason why there should not be absorption systems
with a large N(H{\smc I}) and high metallicity (see more details in text).
The dotted line in the right diagram corresponds to the one-to-one correlation
between the abundances of Zn and Cr that we would observe in the total absence of
dust.
Filled squares correspond to DLA systems with low dust content and open
squares to systems with high dust content. The filled pentagon represents the
host of GRB~010222. The errors in the Pettini sample vary from 1 to 20 per cent for N(H{\smc I}),
and from 0.2 to 0.4 dex for the abundances of Zn and Cr.}
\label{fig:dla}
\end{figure*}

In order to compare the host of GRB~010222 with DLAs detected in QSO
spectra, we have compiled the data published by Pettini and
collaborators (1994, 1997 and 1999).  In Fig.~\ref{fig:dla} we plot
the column density of neutral hydrogen, N(H{\smc I}), versus the Zn
abundance [Zn/H] (i.e. the metallicity) and the abundance of Cr$^+$
(indicator of the dust content) versus that of Zn$^+$. Clearly, the 
host of GRB~010222 is exceptional in a number of ways.
First, the low value of the metallicity
is what one would expect from the extrapolation of the apparent
correlation between N(H{\smc I}) and the metallicity.
There is some controversy regarding the reality of such correlation.
Dust bias in DLA selection is
one of the possible explanations for the absence of high column density, metal rich
absorbers. A more plausible explanation is a cross-sectional bias against 
intercepting the center of galaxies. For a more detailed discussion we refer the reader to Ellison et al. 2001
(and references therein). Second, the abundances of Zn and Cr are the
lowest in the whole sample. 
The correlation of Zn and Cr abundances is caused by the fact that they
are both Fe-peak elements and are produced by similar nucleosynthetic
reaction pathways.  In Galactic stars, Zn and Cr actually track each other
very closely, hence any
departure from [Cr/Zn]$=$0 indicates depletion onto dust. If there was no
depletion on to dust we would expect a one-to-one ratio in Fig~\ref{fig:dla}. Instead, we observe a scatter of
points below this line which is due to the presence of dust.

\begin{table*}
\hspace{-1cm}
\caption{Abundances and dust content in the host galaxy of GRB~010222. The
column densities of Zn and Cr are obtained by averaging the results for the individual
transitions Zn{\smc II} 2026,2062 \AA\AA\ and Cr{\smc II} 2056,2062 \AA\AA\ respectively. The column density of 
Mn and Fe has been obtained from a single transition, Mn{\smc II} 2594 \AA\ and Fe{\smc II} 1608 \AA. 
We assume that the abundance of each element can be approximated by the abundance of its first ionised
ion, since the second ionisation potential is in all cases greater than 13.6 eV.
For comparison, we list as well the range of abundances of these elements through different lines of sight in our Galaxy (see
Savage \& Sembach 1996 and references therein).
Solar meteoritic abundances are taken from Savage \& Sembach (1996).}
\label{tab:abund}
\begin{tabular}{lllllllll}
\hline
Ion		& N$_{\rm{col}}$             & [X/H]  	        & [X/Zn]& $\frac{f_{dust}}{f_{gas}}$& [X/H] & [X/H]     &[X/H]     &[X/H] \\
		&  cm$^{-2}$                 & 010222          &   010222	&		    & Halo & Disk+Halo & Warm Disk & Cool Disk\\
\hline
\hline
Zn$^+$	& (1.25\mpm 1.64)\ET{13}    & -2.3\mpm 0.4     &  -	&  -	&	\si 0	& \si 0	& \si 0	& \si 0 \\
Cr$^+$	& (6.21\mpm 2.46)\ET{13}    & -2.6\mpm 0.7     & -0.34	& \si 1.2   & (-0.38,-0.63) & (-0.72,-0.88) & (-1.04,-1.15) & (-2.08,-2.28) \\
Mn$^+$	& $>$(3.53\mpm 2.35)\ET{13} & $>$-2.7\mpm 0.9  & -0.43	& \si 1.7   & (-0.47,-0.72) & (-0.66)       & (-0.85,-0.99) & (-1.32,-1.45) \\
Fe$^+$	& $>$(5.05\mpm 1.13)\ET{14} & $>$-3.5\mpm 0.6  & -1.26	& \si 17    & (-0.58,-0.69) & (-0.80,-1.04) & (-1.19,-1.24) & (-2.09,-2.27) \\
\hline
\hline
\end{tabular}
\end{table*}

\begin{figure*}
\vspace{0cm}  
\hbox{\psfig{figure=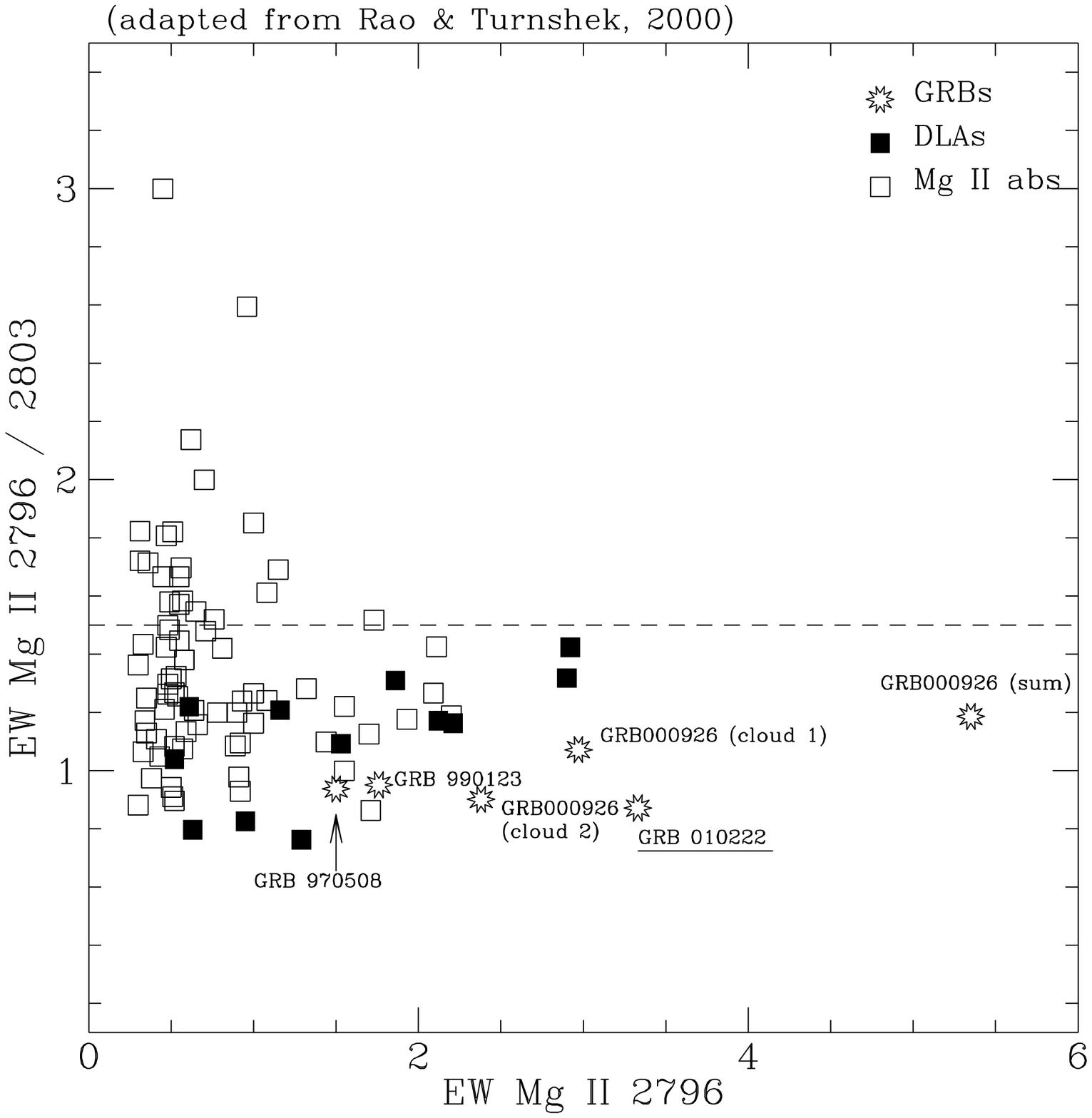,width=8cm} 
\psfig{figure=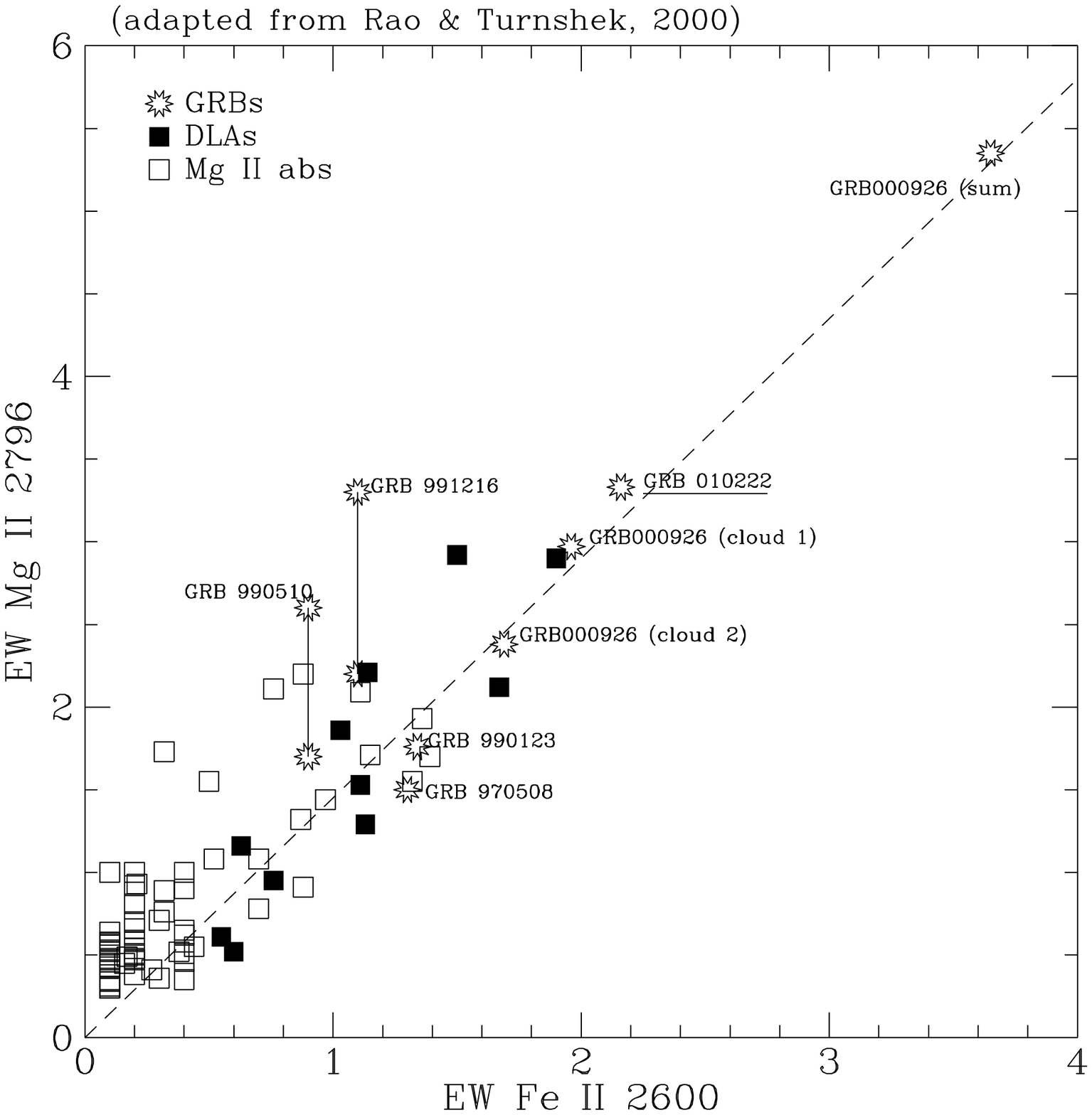,width=8cm}}
\caption{Are all GRB host galaxies damped Lyman-$\alpha$ systems? {\bf
Left:} The ratio of the equivalent width of the two Mg{\smc II} doublet lines is below 1.5 for all
the confirmed DLA systems in the sample of Rao \& Turnshek (2000) {\it and} for those GRB for which
this ratio has been measured (see label in figure).
 {\bf Right:} Another indication that GRB host
galaxies are DLAs: The equivalent width of Mg{\smc II} 2796~\AA\ is correlated to that of
 Fe{\smc II} 2600~\AA\ for the DLA systems of the sample of Rao \& Turnshek (2000). 
The dotted line corresponds to the correlation given in Rao \& Turnshek (2000). 
The GRB host galaxies for
which this information is available follow this correlation.  Moreover, all GRB
hosts have equivalent widths of Fe{\smc II} 2600~\AA\ larger than
0.8~\AA, and of Mg{\smc II} 2796~\AA\ larger than 2~\AA. 
When the individual components of the Mg{\smc II} lines were not be resolved, two
points are drawn in the diagram connected by a vertical line: the
upper value is the total measured EW and the lower value is the EW 
divided by two.}
\label{fig:ratioMgII}
\end{figure*}

\subsection{Other cases of DLAs in host galaxies of GRBs}

This is not the first time that a GRB host
galaxy has been confirmed as a damped Lyman-$\alpha$ system. Jensen et al. (2001)
detected Lyman-$\alpha$ in the afterglow
spectrum of GRB~000301C, and deduced a neutral hydrogen column density
$\log{N(H{\smc I})}$ = 21.2\mpm 0.5 \cm2, qualifying the host galaxy
of GRB~000301C as a DLA.
Another case where the line of Lyman-$\alpha$ is as well detected is the host galaxy of GRB~000926, where the column
density of neutral hydrogen is $\log{N(H{\smc I})}$ \si\ 21.3 \cm2\ (Fynbo et al. 2001).
Although both cases have N(H{\smc I}) lower than in GRB~010222, they are still at the high end of the range found for 
typical DLAs.

It is useful to compare these results with those of Galama \& Wijers (2001). From the analysis of the X-ray spectra
of GRBs afterglows, they find evidence for high column densities (\gapprox 3\ET{21} \cm2) of circumburst material.
The X-ray analysis has the advantage that it measures the column density of
all atoms, irrespective of whether they are in gas or dust. However,
at 1-2 keV rest-frame energies, the X-rays only measure the abundance of elements
whose K edge is at or below that energy (roughly elements with Z$<$10) and therefore one needs to know the
metallicity in order to deduce N(H{\smc I}). The usual practice is to assume solar metallicity which means 
that the real N(H{\smc I}) could be even larger, supporting the idea that GRBs always happen in DLA systems.\footnote{
In principle, when calculating N(H{\smc I}), one should account for any intervening galaxies. Typically, 
those values will be similar to our own galaxy at high latitude, a few times \E{20} \cm2, which means that
if one finds such high values, N(H{\smc I}) $>$\E{22} \cm2, it is most likely due to the host galaxy.}

\begin{figure*}
\vspace{0cm} 
\psfig{figure=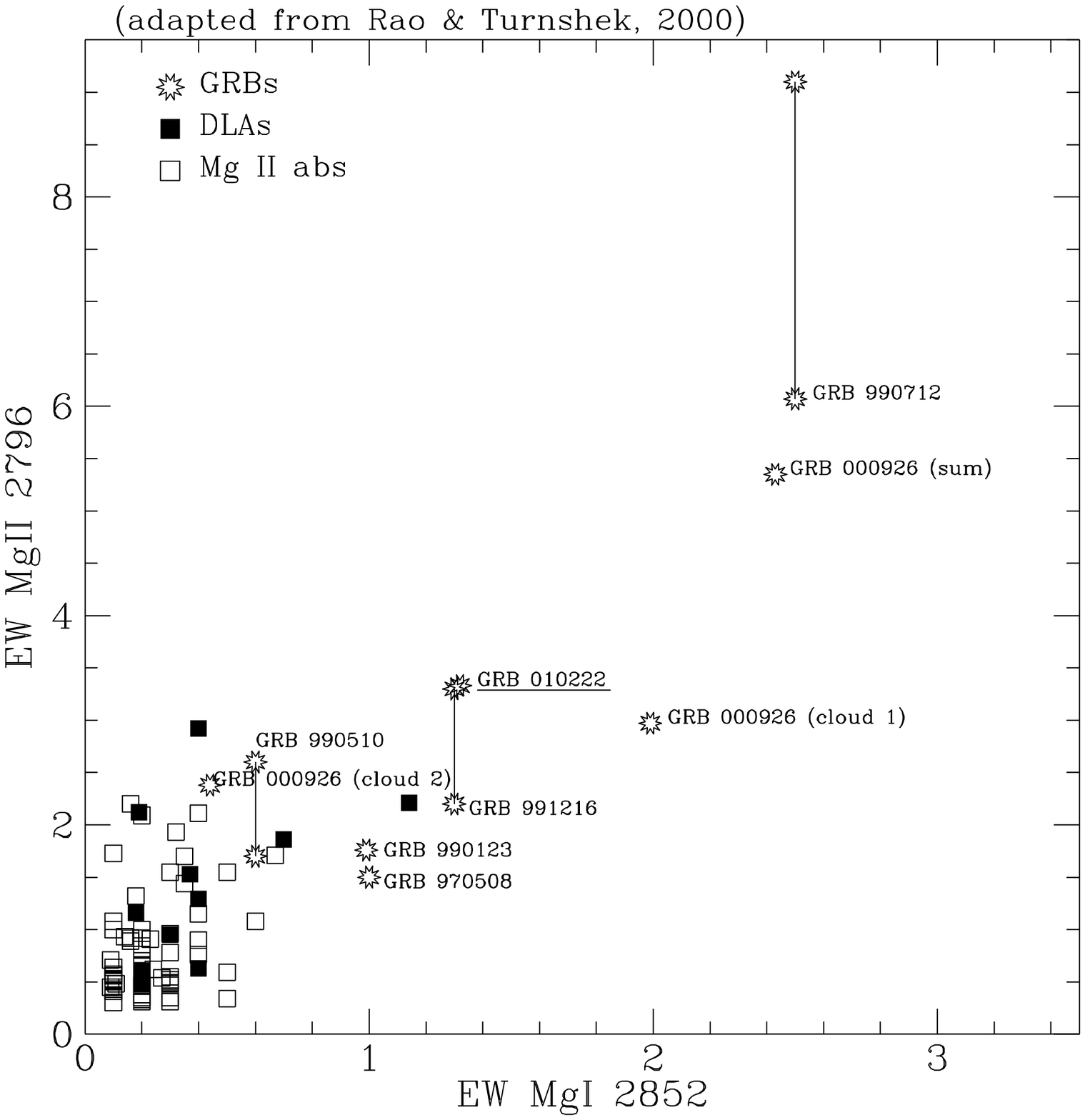,width=8cm}
\caption{Relationship between the equivalent width of Mg{\smc II} 2796 \AA\ and that of Mg{\smc I} 2852 \AA. 
There is a clear trend for both quantities being correlated, that is not evident
if one takes only into account the DLA systems of the Rao \& Turnshek (2000) sample. 
The dots joined with a vertical line correspond to those cases where the Mg II doublet lines
were blended: the upper value is the total measured EW and the lower
value is the EW divided by two.}
\label{fig:fig5}
\end{figure*}

Furthermore, analysis of the equivalent widths of Mg{\smc II} 2796,2803\AA\AA, Mg{\smc I} 2853 \AA\ and
Fe{\smc II} 2600 \AA\ lines provides further support to the idea that the sites of GRB explosions are
DLA systems. We have taken from the literature the EWs of those lines seen in GRB afterglow spectra and added
these points to the diagrams (figs. 24 and 25) presented in Rao \& Turnshek
(2000). The results are shown in Fig.~\ref{fig:ratioMgII}. It is
evident that not only the host of GRB~0102222
would qualify as a DLA on the basis of these diagrams, but also the
hosts of GRB~970508, GRB~990123, GRB~000926, and possibly of GRB~990510
and GRB~991216 (Metzger et al. 1997; Andersen et al. 1999; Kulkarni et al. 1999; Castro et al. 2001; Vreeswijk et al. 2001). 
These results reinforce the idea that all GRB host galaxies are
DLAs. In addition, Rao \& Turnshek (2000) conclude, 
apart from the fact that
all the systems in their sample with EW(Mg{\smc I} 2582) $>$ 0.7 \AA\
are DLAs, that there is no particular correlation
between the EWs of Mg{\smc II} 2796 \AA\ and Mg{\smc I} 2582 \AA\
lines. However, this changes if we add the values for the GRB
mentioned before: there is a clear trend for both EW being proportional to each other (see Fig.~\ref{fig:fig5}).

\section{Discussion}
\label{sec:discussion}

From the analysis of the optical spectrum of the afterglow of GRB~010222, 
we conclude that its host galaxy is a DLA absorption line system, with the strongest column
density of neutral hydrogen and the lowest metallicity ever known, and with low dust
content. 
On the other hand, based on the bright millimetre and sub-millimetre
constant emission (3.74 \mpm 0.53 mJy at 350 GHz), Frail et al. (2001) conclude
that the host galaxy of GRB~010222
is a dusty galaxy with an intense burst of star formation (\si 500
\msunyr)
while in the optical and near-infrared it is a blue sub-luminous galaxy.

There is an apparent contradiction, because Pettini et al. (1999),
based on the lack of metallicity
evolution from z=0.5 to z=3.5,  concluded that
DLAs do not trace the bulk of star-forming galaxies at these redshifts.
Indeed, low redshift imaging studies (e.g. Le Brun et al. 1997) have
shown that DLAs represent a wide range of morphological types.  At higher
redshifts, the low spin temperatures inferred from the 21 cm absorption line studies 
indicate that DLAs rarely represent massive galaxies and may be more
analogous to metal-poor dwarfs (Chengalur \& Kanekar 2000).
On the other hand, the properties (column density,
profile shapes, velocity dispersion and number of components) of Mg{\smc II}
lines seen in the spectra of QSOs seen at z \si 1.5 are very similar to those
observed in $\eta$ Carina nebula. This means that these lines could be
associated with an expansion event, like superbubbles driven by star
formation (Danks 1999). However, Bond et al. (2001), after
comparing the kinematics and redshift evolution of strong Mg{\smc II} absorbers, conclude that 
superwinds and star-forming galaxies can account for a 
substantial fraction of these, but not for DLAs which show different kinematics and no red-shift 
evolution.

Furthermore, DLA
systems having N(H{\smc I}) $> 10^{22}$ \cm2 are very rare. In fact,
to our knowledge, the host of GRB~010222 is the only one known.
One of the explanations often suggested was
that such systems would have a large dust content and therefore would
be missed in optical surveys, but this cannot be the dominant effect because
the survey of radio selected QSOs (and therefore circumvent the
dust issue) by Ellison et al. (2001) did not find any DLA with a large
 N(H{\smc I}). 
Another possible reason proposed by Schaye (2001a) 
is that clouds having such high column densities have a large fraction of molecular 
hydrogen, and such clouds are much smaller and shorter lived than clouds
with low molecular fractions.
Furthermore, at high redshift, there is a
large population of compact galaxies with high star formation rates
and large-scale outflows (called Lyman Break Galaxies, LBG), whose
spectra resemble those of DLA systems seen through the line of sight
towards QSOs. These sight-lines will most likely intersect the outer parts of the
galaxy, while the LBGs will probe the inner parts of the galaxy (for a
detailed discussion see Schaye 2001b).
The cross-section of such clouds will be very small, therefore making the
chance bisection of a background QSO unlikely.  If GRB~010222, and
GRBs in general, are associated with young, violent star formation
regions, that is, with giant molecular clouds, which most likely have large N(H{\smc I}), then the place to
look for high column density DLAs is the spectra of the GRB afterglows,
instead of those of QSOs.  

Another issue of controversy is dust. To explain the fact that 
the optical/UV emission of the host galaxy is faint compared to the sub-milimetre emission,
Frail et al. (2001) propose obscuration by dust.
However, if the system had a high content of dust,
there should be a higher amount of extinction and a larger depletion
of certain elements, in particular Cr, than the one observed (see section~\ref{sec:nature}). 
One possible explanation is that the dust is destroyed by the UV and
X-ray radiation from the GRB and its afterglow. According to Fruchter
et al. (2000), the X-ray emission of the burst can destroy dust (by
grain charging mainly) within the first minutes out to 20 pc. Also, according to
Waxman \& Draine (2000), the UV flash of the burst can sublimate dust
in the first hours after explosion. This would explain the low
extinction and the absence of the 2175 \AA\ feature which is due to
small graphite particles (see also Galama \& Wijers 2001 and
references therein). However, according to Draine (2000) if the GRBs
are associated with a molecular cloud, the UV flash would 
vibrationally excite H$_2$ which would then cause an
absorption at wavelengths \lapprox 1650 \AA\
which would be visible over tens of days. We do not observe such 
feature in our spectrum. 
Moreover, if dust is destroyed by the GRB and its afterglow emission,
the ionization front should have affected the relative abundances
of the different ions for the same element. 
Another factor that we must take into account is that the initial size distribution of dust
particles could be different to that in the diffuse Galactic media,
with a deficiency of small grains and an increase of large ones. This
is what is expected in very dense enviromnents -such as molecular
clouds- where small particles are systematically removed (Kim 1994)
and it is indeed observed in the Small Magellanic Cloud (Weingartner \& Draine 2001,
Welty 2001) and in some Seyfert galaxies (Maiolino et al. 2001).  If
this applies to the environment where the GRBs take place then the
feature due to silicate grains at 9.7 $\mu$m should be absent as well.

A word of caution: the low value of [Zn/H] that we find for GRB~010222 does not necessarily mean 
that its host galaxy is especially metal poor. It could just mean that cold,
dense, giant molecular clouds are particularly depleted. 
Given the fact that the QSOs are unrelated to the absorption-line sites
and the GRBs may not be, it is unclear whether the GRB host galaxies are special or only the GRB environments within hosts.
The low resolution of our spectra does not allow us to disentangle these two cases. 
There is one case, GRB~000926, where Castro et al. (2001) find
that the circumburst material is formed by two different clouds with a
velocity separation of 168 \kms. In these two clouds the relative 
abundances of Zn and Cr are similar to each other, thus indicating that the metallicity and dust 
depletion are characteristic of the host galaxy as a whole,
and ruling out the hypothesis of grain destruction in the vicinity
of the burst. In the case of GRB~010222 there is also some 
substructure with a velocity separation of 106 \kms (Castro et al. 2001).
If these subsystems had similar abundances, then the metallicity and
dust content that we find would be representative of the galaxy as a whole, and
again, dust destruction by the burst itself would be ruled out.

\section{Conclusions}
\label{sec:conclusions}

The observational evidence points out that GRB~010222 has taken place in a damped Lyman-$\alpha$
system having the highest column density of neutral hydrogen and the lowest metallicity 
known up to now, and with low dust content.
This result reinforces the idea that a small cross-section, and not dust, is the
cause of not finding DLA systems having very large N(H{\smc I}). 
If the system
associated with GRB~010222 is not atypical, this means that we may be
missing an important fraction of H{\smc I} at cosmological distances
which in turn may have an impact on $\Omega_{DLA}$ (i.e. the
fraction of neutral gas in DLAs as a function of the closure density),
although the exact contribution will depend on their cross
section, which is unarguably small.

It is not clear whether the dust is destroyed by the GRB itself, or if its size distribution is different, being 
devoided of small particles. In the latter case the
feature due to silicate grains at 9.7 $\mu$m should be absent also. 

Finally, there is strong evidence that other GRB host galaxies (or at least their locations) are as well damped Lyman-$\alpha$ 
systems. Since they seem as well associated with star formation, we conclude that they may
occur in a sub-class of DLAs that {\it is} associated with active star formation. 


\section*{Acknowledgements} 

I.S. and P.M.V. acknowledge support from NWO (post-doctoral grant no. 614.051.003 and Spninoza grant respectivley). 
L.K. is supported by the Royal Academy of Arts and Sciences in The Netherlands (KNAW).
We thank the staff at La Palma who performed the ToO observations.

\bsp 
 
\label{lastpage} 

\end{document}